\documentclass[aps,prl,preprint,groupedaddress]{revtex4-1}
\usepackage{amsmath,amssymb}
\usepackage[dvips]{graphicx}

\begin{document}
\preprint{EPHOU-15-007}
\title{Electroweak Hall Effect of Neutrino and     
  Coronal Heating}
\author{Kenzo Ishikawa and Yutaka Tobita}
\affiliation{Department of Physics, Faculty of Science, Hokkaido
University Sapporo 060-0810, Japan}

\date{\today}
\begin{abstract}
The  inversion of temperature at the solar  corona 
 is hard to understand   from    classical physics,
 and   the coronal
 heating mechanism remains unclear. The heating in the quiet region seems 
 contradicting  with the thermodynamics and is a keen problem for
 physicists.   
A new mechanism for the coronal heating based on the  neutrino radiative
 transition unique  in the
 corona region is studied. The probability  is enormously amplified
 by an electroweak  Chern-Simons form and overlapping waves, and the
 sufficient energy  is transfered.
Thus  the coronal heating is understood from the  quantum effects of the
 solar neutrino. 
\end{abstract}
\maketitle

\section{Introduction} 
  The  experiments  on solar neutrino \cite{daves,koshiba,SK,SNO} proved  that    
  the nuclear fusion  is   the heat
source in the core and the  neutrinos have masses. Now,
there remains a problem   on the temperature of the solar corona.
The sun's temperature is $10^{8}$ K at the core, and $6\times 10^3$ K at
 the solar sphere. That is $10^6-10^{7}$ K at the corona region
  \cite{Grotrian,Edlen}, which  is  higher by $10^3-10^4$  than that at
  the solar surface, despite the fact that the heat source
is in the core. There have been many studies based on the
  electromagnetic interactions, using
  magnetic-acoustic waves, Alfven waves, micro-flare reconnection,
  and others,  which have shown that these regions are not static but
  dynamic of revealing many activities \cite{Golub}. New observation  using satellite 
shows also these  activities. The corona is heated not only in these active
  regions but also in the quiet regions, which seems contradicting with
  the thermodynamics. To find the  heat source  in the quiet
  region is a fundamental physical problem. Hence we focus our study on
  the quiet region in the present paper.

Because the neutrino is produced in the nuclear fusion, about 10\%  
of the initial energy produced at the core is carried by the  neutrino.
Its interaction with matter is quite  weak  of the
  cross section $G_F^2 E_{\nu}m_\text{proton}$ and of a 
  mean free path longer than $ 10^{25}$ m  at the density, 
$n=10^{20}\text{ m}^{-3}$,  and  $E_{\nu}=1$
  MeV. Neutrinos  have been considered  not to interact with the corona.
  It is noted that  these values  
 are  obtained using the transition rate obtained from Fermi's
  golden rule and its equivalent formula of S-matrix  in the relativistic 
field theory, which are valid only if  the initial and final
waves do not overlap \cite{goldberger-watson} and behave like particles.
The neutrino and photon are
waves of large extensions in the corona, and the transition probability
 $P$ gets modified and  has a new term
  $P^{(d)}$ in addition to the standard $T$-linear term 
\begin{eqnarray}
P=\Gamma  {T}+P^{(d)},\label{total probability}
\end{eqnarray}
 where $ {T}$ is the time-interval between the initial and final states 
\cite{Ishikawa-Tobita-ptep,Ishikawa-Tobita-annals,Ishikawa-Tajima-Tobita}.
The term $P^{(d)}$ has origins in the overlap of the waves and consequently is
very different from $\Gamma$ and becomes huge for the neutrino radiative
process in the corona owing to  the tiny photon's
effective mass.  
 $\Gamma $ in the neutrino-photon interactions  is
 ignorable due to Landau-Yang-Gell-Mann theorem
 \cite{landau,yang,gell-mann}, and by a tiny transition magnetic moment
 \cite{Raffelt, Fujikawa-Shrock}, 
but due to $P^{(d)}$   the neutrino radiative transition occurs
 \cite{Ishikawa-Tajima-Tobita}.

Here we consider $P^{(d)}$ of the neutrino radiative process in the corona
region.
The corona has a magnetic field and  free electrons,
which reveals   the Hall effect, which is  expressed  in quantum theory
by a Chern-Simons form of the electromagnetic potential.   
The
 form  is  proportional to ${n_e \over B}{e^2 \over h}$, where $n_e$ is the electron density and $B$ is the
 magnetic field, and agrees with a
 topological invariant. That is   materialized
 as a macroscopic quantum phenomenon  such as the quantized  Hall effect
   in two-dimensional semi-conductors \cite{klitzing,ishikawa}, 
and  is used as the standard of the electric
 resistance.

\section{Induced electroweak Chern-Simons term and  transition
probability}

 In the corona,  
  the  Lamor oscillation in a mean free path  is larger than unity,
  $\omega_B \times {l_{\text{mfp}}^e \over v} >1 $, where  $\omega_B={eB \over
 m_e}$, $l_\text{mfp}^e$ and $v$ is the mean free path and the velocity of the
 electrons. The electrons are expressed
   by Landau levels, which differs from  a weak $B$ expansion
  \cite{Battacharya}.

A system of electrons, photons, and neutrinos in the external magnetic
field in the 3rd-direction are described by the Lagrangian
\begin{align}
&\mathcal L= \mathcal L_0 +{G_F \over \sqrt 2} \bar{\psi}_e(x)
\gamma_{\mu}(1-\gamma_5)\psi_e(x) \bar{\nu}_e(x)
\gamma_{\mu}(1-\gamma_5)\nu_e(x) +e j_{\mu}( A^{\mu}_\text{ext}+A^{\mu}), \nonumber\\
&\mathcal L_0= \bar \nu_i(x)(p^{\mu}\gamma_{\mu}(1-\gamma_5)-m_i) \nu_i(x)+
\bar{\psi}_e(x)(p^{\mu}\gamma_{\mu}-m_e)\psi_e(x)-
\frac{1}{4}F_{\mu\nu}F^{\mu\nu},
\label{lagrangian}
\end{align}
where  the magnetic field is expressed by  $A^{\mu}_\text{ext} $, and $\nu_e(x)$
is the electron neutrino. The
neutral current interaction is symmetric in all flavours and  does not
contribute to  the neutrino radiative transitions and were ignored 
in Eq. $(\ref{lagrangian})$. 
Expanding $\psi_e(x)$ with  eigen functions of including $A^{\mu}_\text{ext} $, and
integrating  them, we   find  the
effective Lagrangian \cite{ishikawa,imos,iaim},
\begin{eqnarray}
\mathcal L_\text{int}=\frac{\nu^{(4)}}{2\pi}\epsilon^{\alpha \beta \gamma} \tilde
 A_{\alpha} \partial_{\beta} \tilde A_{\gamma}  +O({\tilde
 F_{\alpha\beta}}^2);\ \alpha,\beta,\gamma=(0,1,2),
\label{effective-lagrangian}
\end{eqnarray}   
where $\nu^{(4)}= { 2 \pi \hbar n_e \over eB} $
 is the filling factor   of Landau
levels, and
 $\epsilon^{\alpha\beta\gamma}$ is the anti-symmetric tensor,
 and
\begin{align}
&\tilde A_{\alpha}=e A_{\alpha}(x)+{G_F \over \sqrt 2} J_{\alpha}(x),\\
&J_{\alpha}(x)=\bar \nu_e(x)\gamma_{\alpha}(1-\gamma_5)
 \nu_e(x),\\
 &\tilde F_{\alpha\beta}= F_{\alpha\beta}+{G_F \over \sqrt 2}(\partial_{\alpha}
  J_{\beta}-\partial_{\beta} J_{\alpha}).
\end{align}
 $\nu_e(x)$ is the superposition  $\nu_e(x)=\sum_i U_{ei}\nu_i(x)$ of three  
mass eigenstates $\nu_i(x);\ i=1-3$ and  a  mixing matrix  $U$. It follows 
that  
\begin{eqnarray}
& &J_{\alpha}(x)= g_{ij} \nu_i(x) \gamma_{\alpha}(1-\gamma_5)\nu_j, \\
& &g_{ij}=U_{ei}^{*}U_{ej}. \nonumber
\end{eqnarray}
Thus Eq. \eqref{effective-lagrangian}
leads   a neutrino radiative transition, which has the following unusual
properties:  that is  Lorentz non-invariant; the strength is
proportional to  $e G_F \nu^{(4)}$.  The coupling strength is a topological invariant which satisfies 
a low energy theorem and   remains the same in systems of
 disorders at finite-temperature
 \cite{ishikawa,imos,iaim}.

A radiative transition  $\nu_i \rightarrow
\gamma+\nu_j$  takes place,   
since the neutrino mass difference is larger  than  the photon's effective mass
 determined by the plasma frequency  $m_{\gamma,\text{eff}}=\hbar
 \omega_p$.
The event that $\gamma$ is detected or interacts with others at $T$ 
and $\nu_j$ escapes, is studied.  The LSZ formula
\cite{LSZ,Low} is extended to an
S-matrix, $S[ T]$ of satisfying  this  boundary condition  
\cite{Ishikawa-Tobita-ptep,Ishikawa-Tobita-annals,Ishikawa-Tajima-Tobita}
 at the finite-time interval. Because $[S[ T],H_0] \neq 0$, $S[ T]$
couples with the final states of  continuous  spectrum of the kinetic
energy different  from the initial kinetic energy, which is caused by 
the overlap of the waves. Thus the space time symmetry of the free Lagrangian
such as the conservation law of the kinetic energy and manifest Lorenz
invariance are partly broken in $P^{(d)}$. 
Being non-invariant, $P^{(d)}$ can  be much larger in magnitude
than the invariant as  in the example;  $
|\vec p_{\nu}|^2 \gg E_{\nu}^2-(\vec p_{\nu})^2=m_{\nu}^2$, 
at the  high energy.
$\Gamma$ is Lorentz invariant and is proportional to $m_{\nu}^4 $, 
which is negligibly small for the neutrino \cite {particle-data,Tritium,WMAP-neutrino},
whereas $P^{(d)}$ is proportional to a lower power in $m_{\nu}$ of   much larger magnitude.

The amplitude  
 is written as $\mathcal M=\int d^4x \, \langle {\gamma},{\nu_j}
 |(-L_\text{int}(x))| \nu_i \rangle$, for     a neutrino   prepared at a time 
 ${T}_{\nu_i}=0$,  and  a photon  interacting  at a space-time position 
$({T}_{\gamma},\vec{{X}}_{\gamma})$ and an unmeasured-neutrino, which are
expressed in the form   
$|\nu_i \rangle=   | {\vec p}_{\nu_i},{T}_{\nu_i}=0  \rangle,\ 
|\nu_j ,\gamma \rangle=   |\nu_j,{\vec p}_{\nu_j};\gamma,{\vec
 p}_{\gamma},\vec{{X}}_{\gamma},{T}_{\gamma}          \rangle$,
and  the time $t$ is
 integrated in the region $0 \leq t \leq T_{\gamma}$. 
 The size of photon wave function,
${\sigma_{\gamma}}$,  is estimated later. 
After the straightforward but tedious calculations, the details of which
 were given in Refs. \cite{Ishikawa-Tobita-ptep,Ishikawa-Tobita-annals}, we have the total
probability in the form  
\begin{align}
\label{probability-3}
P=N_2\int \frac{d^3 p_{\gamma}}{(2\pi)^3 E_{\gamma}}
{(\tilde p_{\nu_i} \cdot \tilde p_{\gamma})(\tilde p_{\nu_i} \cdot
 \tilde p_{\gamma}-{\tilde p_{\gamma}}^2 )}
 \left[\tilde g(\omega_{\gamma}, {T}) 
 +G_0 \right],
\end{align}
where $\tilde p=(p_0,p_1,p_2)$,  $N_2 = 8 {T} ({\nu^{(4)} e G_F g_{ij}\over 2\pi})^2   {\hbar
\sigma_{\gamma} \over \epsilon_0 E_{\nu_i}}$, ${L} = c{T},\ {T}={T}_{\gamma}$ is
the length of decay region. The function $\tilde g(\omega_{\gamma},{T})$, which is  given in
Refs. \cite{Ishikawa-Tobita-ptep,Ishikawa-Tobita-annals}, is characterized by a
phase factor $e^{i\omega_{\gamma}(t_1-t_2)}$ of the correlation function
of the angular velocity 
$\omega_{\gamma}=\hbar \frac{m_{\gamma, \text{eff}}^2 c^4}{2E_{\gamma}}$, 
shows $P^{(d)}(\gamma)$,   and  $G_0$ shows  $\Gamma  T$ and 
  is negligible now. The phase space for $P^{(d)}(\gamma)$ is different from
  that of $\Gamma  T$, and is derived from the causality condition.      
It follows for the general cases of two  neutrino
  flavour and an angle $\Theta_{{\vec B},{\vec p}_{\nu_1}}$ between the magnetic field and ${\vec
  p}_{\nu_1}$ that
\begin{eqnarray}
& &P^{(d)}(\gamma) \approx  P^{(d)}_\text{asym}(\gamma) {T \over T_0},\ T_0={1
\over \omega_{\gamma}};\ \omega_{\gamma} T < 1\label{general-probability1},\\
& &P^{(d)}_\text{asym}(\gamma)= \eta {\alpha \over 5\pi} 
\left(\frac{G_F}{ (c\hbar)^3}E_{\nu_1}^2\right)^2
  \frac{\delta m^2_\nu}{m_{\gamma,\text{eff}}^2}
\left({\nu^{(4)} \over 2\pi}\right)^2
 \sigma_{\gamma} \label{general-probability2};\ \omega_{\gamma} T \geq 1,\\      
& & \eta=\cos^2 {\Theta_{{\vec B},{\vec p}_{\nu_1}}} \cos^2 \theta_{12}, \nonumber
\end{eqnarray}
\begin{figure}[t]
\includegraphics[scale=0.45,angle=-90]{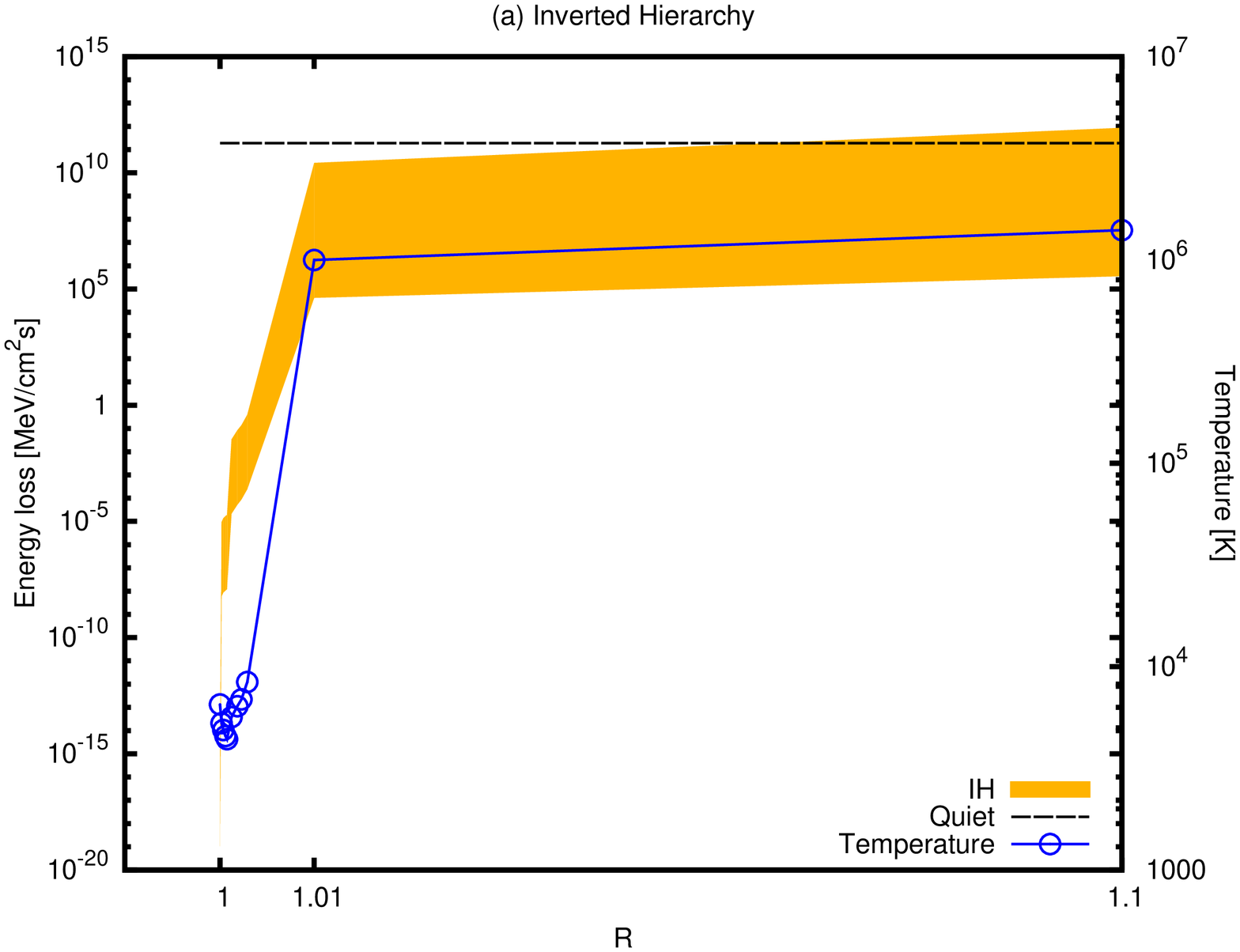}
\includegraphics[scale=0.45,angle=-90]{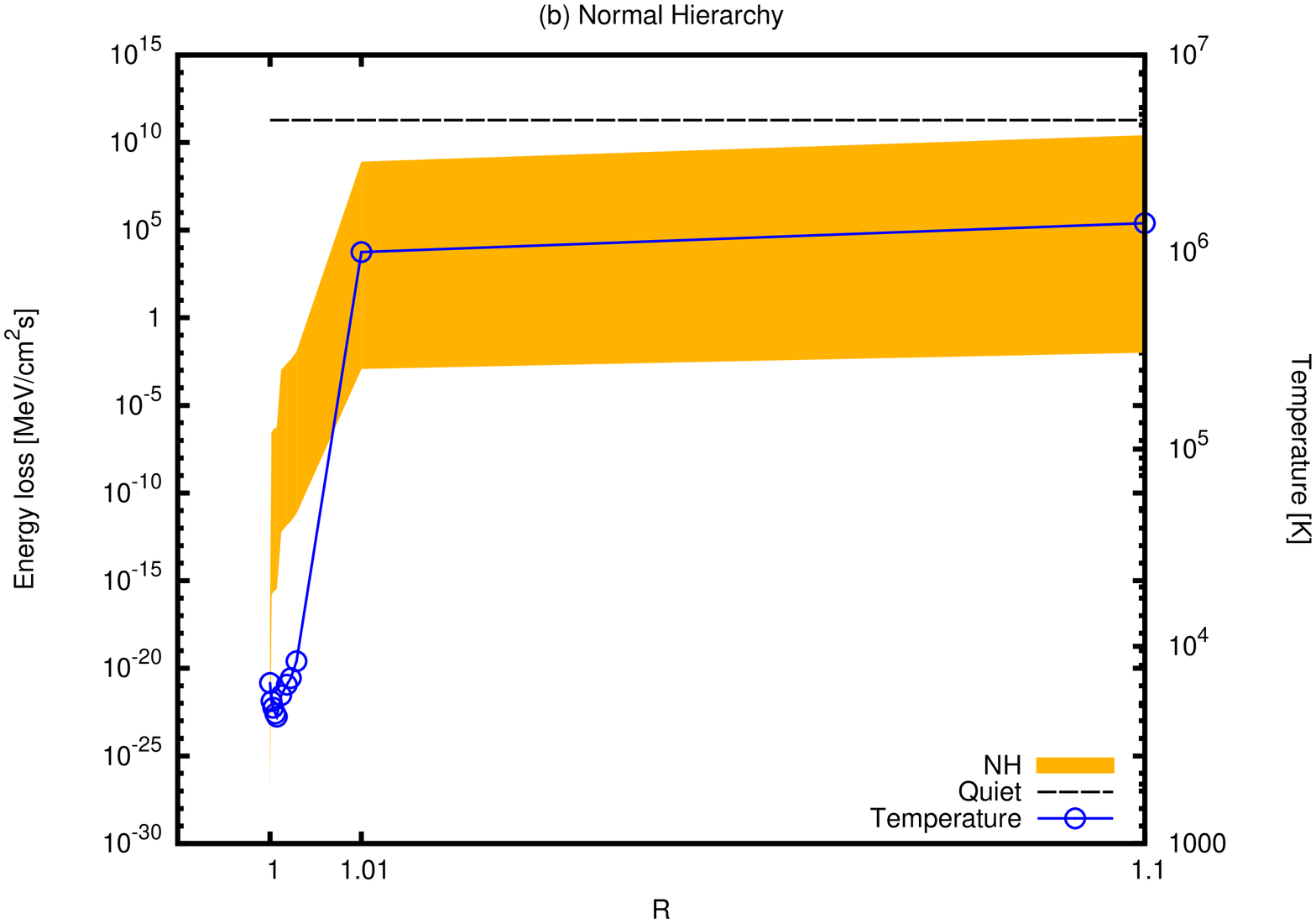} 
\caption{$R$ dependence of the transferred energy to the corona 
 [MeV/cm${}^2$s] is compared with the energy necessary to heat the
 corona at $R=1.1 R_\odot$. Temperature rising steeply at around $R=1.01
 R_\odot$ is also shown. (a) is for the inverted mass
 hierarchy and supplies  the sufficient energy, but    
(b)  for the normal mass hierarchy  does not supply the sufficient energy.}
 \label{fig:EL}
\end{figure}
where   $\theta_{12}$  is a mixing angle between the flavour and
mass eigenstates, and  the heaver neutrino corresponds to the electron
neutrino if $\theta_{12}=0$.
For the three neutrino cases, $\eta$ becomes a more complicated
expression of  an essentially the equivalent result. 

The state vector of revealing 
 the constant probability, $P^{(d)}(\gamma)$, 
 in the vacuum,    is a  superposition of the parent and daughters 
of almost  time independent weight, and is  like a stationary state.  In 
the environment of  many atoms or molecules, on the other
 hand, the produced photon interacts with them  and loses its energy
 easily. Consequently, the probability 
of the neutrino to lose the energy 
is given by the product $P^{(d)}(\gamma) \times P_{\gamma}$, where $P_{\gamma}$ 
is governed by QED and is order
unity.  Hence, the 
energy is transfered, and the
environment gains the energy and is heated. It must be noted that the 
diffractive probability for the neutrino is determined by $\omega_{\nu}$ 
and  is much smaller than that of the photon, in the situation $m_{\nu} \gg
 m_{\gamma}$  of the paper,  and is ignored.

\section{Coronal heating and solar wind 
acceleration }

Equations $(\ref{probability-3})$, $(\ref{general-probability1})$, and 
$(\ref{general-probability2})$ are applied to the solar
corona. Parameters are taken from Ref. \cite{allen} and  
  the filling factor $\nu^{(4)} $, and the wave packet size  $\sigma_{\gamma} =  
(l_\text{mfp}^{c})^2 $, where  $l_\text{mfp}^{c}$ is the mean free
path  of the proton or electron are computed and used.   
The photon's effective mass $m_{\gamma,\text{eff}}$ becomes  smaller
than the neutrino mass
squared difference.    
The function $\tilde{g}(\omega_{\gamma},T)$  at  the non-asymptotic region is
 used. Using these values, we have
\begin{eqnarray}
P^{(d)}(\gamma) \approx 10^{-3};\ E_{\nu}=10 \text{ MeV},
\label{solar-reduction}
\end{eqnarray}
the total
energy transferred to the corona 
\begin{eqnarray}
& &E^\text{transfer}=  P^{(d)}(\gamma) \times N_{\nu}\approx  5 \times 10^5 \text{ erg}
\,\text{cm}^{-2} \text{s}^{-1},
\end{eqnarray}
where $N_{\nu}$ is the energy flux of the neutrino. In Fig. \ref{fig:EL}, 
$E^\text{transfer} $ is given as a function of the radius 
$R/{R_\odot}, R_\odot=\text{the slar radius}$, and compared with the estimated energy at 
$R=1.1 R_\odot$ \cite{withbroe}, and the 
temperature. The mass differences are known but the absolute masses and
mass hierarchy are unknown. For the inverted mass hierarchy,  
the neutrino mass-squared differences  $\delta m_{\nu}^2=2.52 \times 10^{-3}\text{ eV}^2(\text{IH})$ 
and for the normal  mass hierarchy $\delta m_{\nu}^2=7.53\times 10^{-5}\text{ eV}^2(\text{NH})$ 
are substituted and the absolute mass are varied.   
The energy for IH  is in accord with  the 
observation, however  the maximum value for NH is
$1/{100}$ of   the observation.  The normal
mass hierarchy would be  rejected, even though there are  ambiguities on the
magnetic field, the electron density, and the temperature.
 Because the electron
density  decreases steeply,  
the probability $P^{(d)}(\gamma)$ becomes the value Eq. $(\ref{solar-reduction})$
rapidly in the corona region, and the temperature rises steeply, which is in
agreement with the observation. Thus the  heating of the quiet corona is
understood from the neutrino radiative transitions. 

In a corona  hole, the electron density  is low and 
the magnetic field is high, and  $P^{(d)}$ is not large 
at the height of the transition region. However the effect becomes
stronger because the magnetic
field  is parallel to the neutrino
up to a high altitude, and an   acceleration of the solar wind
becomes higher, which   is in
accord  with the observation. A recent observation of a density
modulation of the plasma wave \cite{s-wind} also agrees with the extremely slow
angular velocity $(\omega_{\gamma})^{-1}\approx 10^2  $ s in this region.


The probability $P^{(d)}(\gamma)$ varies following the change 
of $B$, and influences the  solar constant  and other related phenomena.
    $B$  is large in the core of 
the sun spot, the rate of energy loss is correlated with
the sun spot number.  
A  correlation between the small variation of the solar constant of the
order of Eq. $(\ref{solar-reduction} ) $ or slightly smaller value and
 the sun spot number observed in    recent measurements  appears.   

In the active region  where  the present conditions hold,  the
diffractive probability is important as well.

\subsection{Earth Ionosphere and radiation belt }

Ionosphere in the earth  is also a plasma of low density and  weak 
magnetic field, so is affected by the electroweak Hall effect. 
Substituting values $10^{-5}$ T and  $10^{11}\text{ m}^{-3}$ for the magnetic
field and electron density, we have  
\begin{eqnarray}
P^{(d)}(\gamma)=10^{-7},
\end{eqnarray}  
which is much smaller than the value at the solar corona
Eq. $(\ref{solar-reduction} ) $. The  neutrino flux is lower
by  a factor $10^{-6}$ due to the large distance, and the  energy 
released from the neutrino to atoms becomes small. On the other hand,
the temperature of lights from the sun is $5800 \text {K}$ and is higher
than that in the ionosphere   $ 1000$ K, which may be caused mainly 
by photochemical reactions. It may be hard to see
the temperature variation caused by the neutrino in the
ionosphere. Nevertheless the diagonal component 
 of the interaction Lagrangian in the neutrino flavour,  causes the neutrino 
current to induce the electric  or magnetic fields.
Such time-dependent  variation of the magnetic field around $20 \text
{ nT} $ observed at the earth surface may
be connected with  the  solar neutrino through the electroweak Hall effects in the earth.
 These will be presented in a
forthcoming work.


 
\section{Summary and implications.} 

The new coronal heating mechanism is based on the following: 
\textbf{(1)}  massive neutrino, \textbf{(2)}  electroweak Hall effect of the dilute plasma in 
 the magnetic field  expressed by  the effective
interaction Eq. ($\ref{effective-lagrangian}$), and 
\textbf{(3)} the diffractive component of the
transition probability $P^{(d)}(\gamma)$.   
 They lead  
the large coupling strength, ${\nu^{(4)} \over 2\pi}$, the tiny 
photon's effective mass $m_{\gamma,\text{eff}}=\hbar \omega_p$, and the large wave packet
$\sigma_{\gamma}$, and  make $P^{(d)}(\gamma)$ to be $10^{-3}-10^{-4}$.
Due to $P^{(d)}(\gamma)$, the neutrino which arrives the corona region
maintaining  the initial
energy, decays and loses the energy. Its probability  $P^{(d)}(\gamma)\times P_{\gamma}$ agrees 
with $P^{(d)}(\gamma)$, because   $P_{\gamma}\approx 1$.   Hence the average
energy  
$P^{(d)} (\gamma)E_{\nu}/2$, which is in accord with the observation
given in Fig. \ref{fig:EL}, is 
transfered  to the corona gas.  
$P^{(d)}(\gamma)$ was enhanced enormously in the corona region and
the  temperature rises steeply.  The neutrinos gives  the heat to the
solar corona, and a possible electromagnetic effect to the earth through ionosphere.

The probability $P^{(d)}(\nu)$
for the neutrino  is determined by the neutrino mass and
$\sigma_{\nu}$, which are assumed $m_{\nu} \approx 10^{-3} \text{ eV}$ and  
$\sigma_{\nu} \approx {10^{-28} \text{ m}}^2$, and is  much smaller
than those  of the photon. Hence 
a reduction of the neutrino flux is  negligible unless $m_{\nu} \approx
0$. 
The neutrino oscillation experiments measure the $T$-dependent
variations of the neutrino flux derived from  the components $\Gamma
 T$.  The
large $P^{(d)}(\gamma)$ in the dilute electron gas in the magnetic field  
is not in contradiction  with existing neutrino phenomena and experiments. 
A correlation of the neutrino flux with 
the sun  spot 
number  may be too small to measure  using the current ground
detector. 

It would be worthwhile to test  the present mechanism using ground experiments
with nuclear reactors,  high energy accelerators, or others.  

A new quantum phenomenon of the neutrinos radiative decays  was derived 
and shown to give the heat source to the solar corona, and other effects
in the earth ionosphere.  Thus the neutrino produced in the core 
gives the energy   into  the solar corona. The
mechanism for the former is the nuclear fusion and that for the latter
is  the electroweak Hall effect and the diffractive probability.

\begin{acknowledgments}
This work was partially supported
by a Grant-in-Aid for Scientific Research(Grant No. 24340043) provided
by the Ministry of Education,
Science, Sports and Culture, Japan.
One of the authors (K.I) thanks Drs. Takashi Kobayashi, Tsuyoshi Nakaya,
and Fumihiko Suekane  for 
useful discussions on 
the  neutrino experiments, Drs. Shoji Asai, Toshinori Mori, Sakue Yamada, Tomio Kobayashi,
Makoto Minowa, Kazuo Sorai, Mitsuteru Sato, Toshiki Tajima, and Shigeto Watanabe for useful discussions on interferences and
solar and earth systems. 
\end{acknowledgments}

{}
\end{document}